# The quantal algebra and abstract equations of motion

Samir Lipovaca
slipovaca@aol.com

*Abstract*: The quantal algebra combines classical and quantum mechanics into an abstract structurally unified structure. The structure uses two products: one symmetric and one anti-symmetric. The local structure of spacetime is contained in the quantal algebra without having been postulated. We will introduce an abstract derivation concept and generalize classical and quantum mechanics equations of motion to abstract equations of motion in which the anti-symmetric product of the quantal algebra plays a key role. We will express the defining identities of the quantal algebra in terms of the abstract derivation. In this form the first defining identity (the Jacobi identity) is analogous in form to the Bianchi identity in general relativity which is one set of gravitational field equations for the curvature tensor. The identity for the unit element is equivalent to the important property that the metric is covariantly constant. Similarly, the Jacobi identity is analogous in form to the homogeneous Maxwell equations. The anti-symmetric product of the quantal algebra is reflected in the antisymmetry of the electromagnetic tensor.

## Introduction

The root cause of the conflict between quantum mechanics and relativity originates from the difference in the underlying Lie groups and Lie algebras. In quantum mechanics we deal with unitary groups. In relativity we deal with orthogonal group. The quantal algebra [1] is a natural abstract structure of quantum theory. The algebra of quantions is $SO(2,4)$ concrete realization of the quantal algebra. This unique mathematical structure structurally extends non relativistic quantum mechanics to a relativistic theory by generalizing its underlying number system (the field of complex numbers). It contains both quantum mechanics and relativity in exactly four dimensions. Since the translation group does not appear in the algebra of quantions, the space is intrinsic Reimannian. Therefore, the algebra of quantions structurally unifies quantum mechanics and general relativity.

The unnecessary division property of complex numbers was the main road block in uncovering the relativity structure. Null space-time intervals do not have an inverse in the algebra of quantions. Imposing the unnecessary division property removes relativity from the algebra of quantions, forcing us back at using complex numbers. This algebra is the only algebra that is able to lift a degeneracy of complex numbers and has different algebraic and geometric norms. The algebraic norm of complex numbers is defined as

$$A(z) = zz^*$$

where $*$ is complex conjugation. We can expand $A(z)$ in terms of the components of

$z = x + iy$ and obtain

$$A(z) = x^2 + y^2 = M(z)$$

where $M(z)$ is a geometric concept (the trivial Euclidian metric in two dimensions). For complex numbers the algebraic and geometric norms are equal. If this degeneracy is lifted, it will lead uniquely to the algebra of quantions. The two norms of quantions have a remarkable physics interpretation. The algebraic property of quantions is related to standard quantum mechanics. The geometric property is related to relativity. In other words, $A$ maps quantions to complex numbers and non-relativistic quantum mechanics, while $M$ maps quantions into Minkowski four vectors and relativity. Thus quantions are mixed relativity and quantum mechanics objects. In quantionic physics, Born interpretation of the function $\rho = \psi^* \psi$ as a probability density is naturally generalized and replaced by Zovko's interpretation which leads to Dirac and Schrodinger equations.

After these introductory remarks about quantions as a concrete realization of the quantal algebra, let us outline the paper. First, we give a brief overview of the quantal algebra. Then the Hamiltonian as a crucial concept of both classical and quantum mechanics is presented. Based on this fundamental importance of the Hamiltonian in both mechanics, we introduce an abstract derivation concept and generalize classical and quantum mechanics equations of motion to abstract equations of motion in which the anti-symmetric product of the quantal algebra plays a key role. We express the defining identities of the quantal algebra in terms of the abstract derivation. In this form certain concepts of the gauge theories (the algebraic identities and the homogeneous differential equations) are readily apparent. We emphasize these concepts for the gravitational and electromagnetic fields. Last, we finish with Discussion and Conclusions.

## Quantal algebra

A quantal algebra $\{O, \sigma, \alpha, e\}$ is a real abstract two-product algebra with a unit. Denoting the elements of the algebra by small letters, $f, g, h, ... \in O$, where $O$ is the space of observables, the two products, denoted by $\sigma$ (symmetric) and $\alpha$ (anti-symmetric) are Jordan and Lie products respectively. Following are the defining identities of the quantal algebra:

the Jacobi identity, $(f \alpha g) \alpha h + (g \alpha h) \alpha f + (h \alpha f) \alpha g = 0$,

the Leibnitz identity, $g \alpha (f \sigma h) = (g \alpha f) \sigma h + f \sigma (g \alpha h)$,

and the Petersen identity, $(f \sigma g) \sigma h - f \sigma (g \sigma h) = a\, g \alpha (h \alpha f)$.

The identities for the unit element $e \in O$ are

$$e \, \sigma f = f,$$

$$e \, \alpha f = 0.$$

The constant $a$ is related to Planck's constant. For quantum mechanics it may have any positive value. It may be normalized to unity by re-scaling the product $\alpha$. For classical mechanics, we have $a = 0$.

### Hamiltonian in classical and quantum mechanics

Both classical and quantum mechanics specify how the state of a system evolves with time. In classical mechanics, like any other property of the system, its total energy is determined by its state. It is a function of the position and momentum coordinates of the particles comprising the system. This function is known as the Hamiltonian function $H$ of the system. The Hamiltonian function dictates how the state of a classical system evolves through time. The fundamental equations of classical mechanics for a system of $f$ degrees of freedom, written in the so-called "canonical" form are,

$$\dot{p}_k = -\frac{\partial H}{\partial q_k}, \; \dot{q}_k = \frac{\partial H}{\partial p_k}, \; k = 1, 2, \ldots, f. \tag{1}$$

When $H$ does not depend explicitly on the time, the energy equation $H(p, q) = W$, where $W$ the total energy, is a constant, follows at once. The canonical equations (1), if expressed in terms of the Poisson brackets, become

$$\dot{p}_k = [H \; p_k], \quad \dot{q}_k = [H \; q_k] \tag{2}$$

where the Poisson bracket of $x$ and $y$ is defined as

$$[x \; y] = \sum_{k=1}^{f} \left( \frac{\partial x}{\partial q_k} \frac{\partial y}{\partial p_k} - \frac{\partial y}{\partial q_k} \frac{\partial x}{\partial p_k} \right). \tag{3}$$

Like classical mechanics, quantum theory tells us how the state of a system evolves with time. The key role in the equation governing this evolution is played by an operator rather than by the Hamiltonian function, according to the general principle that, in quantum mechanics, operators represent physical quantities. As in the classical case, the quantity in question is the total energy of the system. It is represented in quantum theory by a Hermitian operator $H$ which we call the Hamiltonian operator for the system. Dirac [2] took the point of view that the Hamiltonian is really very important for quantum theory. In fact, without using Hamiltonian methods one cannot solve some of the simplest problems in quantum theory, for example the problem of getting the Balmer formula for hydrogen, which was the very beginning of quantum mechanics. It has been found by Dirac that in the quantum theory the expression $\frac{2\pi i}{h}(xy - yx) = \frac{2\pi i}{h}[x, y]$, where $h$ is the Planck constant, is the analogue of the Poisson bracket (3) in classical mechanics. The simplest assumption is to take over the equations (2) formally into the quantum theory, replacing the Poisson brackets by their quantum analogues. Therefore it is assumed the equations of motion in the quantum theory to be

$$\dot{p}_k = \frac{2\pi i}{h}[H, p_k],$$

$$\dot{q}_k = \frac{2\pi i}{h}[H, q_k].$$
(4)

We see in both classical and quantum mechanics, the Hamiltonian is a crucial concept.

### Abstract equations of motion

A quantal algebra with $a = 0$ represents the abstract structure of classical mechanics. The phase space functions are concrete observables and their ordinary product and Poisson bracket are the algorithms for the abstract products $\sigma$ and $\alpha$:

$$f \sigma g \stackrel{def}{=} fg,$$

$$f \alpha g \stackrel{def}{=} \sum_{k=1}^{f} \left( \frac{\partial f}{\partial p_k} \frac{\partial g}{\partial q_k} - \frac{\partial g}{\partial p_k} \frac{\partial f}{\partial q_k} \right).$$
(5)

Similarly, a quantal algebra with $a > 0$ represents the abstract structure of quantum mechanics. Hermitian matrices are concrete quantum mechanical observables and half the Hermitian commutator and half the Hermitian anti-commutator are the algorithms for the abstract products $\alpha$ and $\sigma$:

$$f \alpha g \stackrel{def}{=} \frac{1}{2i}(fg - gf),$$

$$f \sigma g \stackrel{def}{=} \frac{1}{2}(fg + gf).$$
(6)

Based on the equations (2) and (4) it seems reasonable to assume the abstract equations of motion to be

$$\Delta_f g = g \alpha f \qquad (7)$$

where $\Delta_f$ is the abstract derivation associated with $f$ in a sense if $f$ represents a Hamiltonian then a concrete realization of $\Delta_f$ is the time derivative $\frac{d}{dt}$. The following observations seem to make (7) plausible. For $a = 0$ and $\Delta_f = \frac{d}{dt}$, using the algorithm (5) for the abstract product $\alpha$ the abstract equations of motion become the canonical equations (2):

$$\frac{d}{dt}g = \dot{g} = [fg].$$

Here $g$ stands for $p_k, q_k$ and $f$ stands for the Hamiltonian function $H$. Similarly, for $a > 0$ and $\Delta_f = \frac{d}{dt}$, using the algorithm (6) for $\alpha$, the abstract equations of motion become the equations of motion in the quantum theory for $g = p_k, q_k$ and $f = \frac{4\pi}{h}H$:

$$\frac{d}{dt}g = \dot{g} = \frac{1}{2i}[g, \frac{4\pi}{h}H] = \frac{2\pi}{ih}[g, H] = \frac{2\pi i}{h}[H, g].$$

When $f = g$ from equation (7) follows $\Delta_f f = 0$ since $\alpha$ is an anti-symmetric product. Thus $f$ is a constant of motion with respect to $\Delta_f$. In both mechanics this means $\dot{H} = 0$. If we express the defining identities of the quantal algebra in terms of the abstract derivation we obtain:

the Jacobi identity, $\Delta_h \Delta_g f + \Delta_f \Delta_h g + \Delta_g \Delta_f h = 0$,

the Leibnitz identity, $\Delta_{(f \sigma h)} g = (\Delta_f g) \sigma h + f \sigma \Delta_h g$, \qquad (8)

and the Petersen identity, $(f \sigma g) \sigma h - f \sigma (g \sigma h) = a \Delta_{\Delta_f h} g$.

The identity for the unit element is $\Delta_f e = -\Delta_e f = 0$. The unit element behaves like a constant with respect to the abstract derivation associated with any element of the algebra. Similarly, any element behaves like a constant with respect to the abstract derivation associated with the unit element. Remarkably, in this form of the quantal algebra certain concepts of the gauge theories (the algebraic identities and the homogeneous differential equations) are readily apparent. Let us emphasize these concepts for the gravitational and electromagnetic fields.

## The gravitational field

From the Jacobi identity in (8) we can subtract another Jacobi identity in which $h$ and $g$ are interchanged

$$\Delta_g \Delta_h f + \Delta_f \Delta_g h + \Delta_h \Delta_f g = 0$$

to obtain

$$(\Delta_h \Delta_g - \Delta_g \Delta_h)f + (\Delta_f \Delta_h - \Delta_h \Delta_f)g + (\Delta_g \Delta_f - \Delta_f \Delta_g)h = 0. \qquad (9)$$

If $f$ represents a covariant vector $V_f$ and $\Delta_h$ a covariant derivative $\nabla_h$ then $(\Delta_h \Delta_g - \Delta_g \Delta_h)f$ gives

$$\nabla_h \nabla_g V_f - \nabla_g \nabla_h V_f = V_{f;g;h} - V_{f;h;g} = V_c R^c_{fgh}$$

where $R^c_{fhg}$ is the Riemann-Christoffel tensor or the curvature tensor. Similarly, we can replace other two terms in (9) with covariant vectors and covariant derivatives to obtain

$$V_c R^c_{fgh} + V_c R^c_{hfg} + V_c R^c_{ghf} = 0.$$

The factor $V_c$ occurs throughout this equation and may be cancelled out. We are left with the first Bianchi identity

$$R^c_{fgh} + R^c_{hfg} + R^c_{ghf} = 0.$$

Let us suppose now $f, g,$ and $h$ represent $\nabla_f V_a, \nabla_g V_a, \nabla_h V_a$ respectively, where $V_a$ is a covariant vector. Then

$$(\Delta_h \Delta_g - \Delta_g \Delta_h)f = (\nabla_h \nabla_g - \nabla_g \nabla_h)\nabla_f V_a = (\nabla_h \nabla_g - \nabla_g \nabla_h)V_{a;f} =$$
$$= V_{a;f;g;h} - V_{a;f;h;g} = V_{c;f} R^c_{agh} + V_{a;c} R^c_{fgh}.$$

Similarly, for the second and third terms in (9) we obtain

$$(\Delta_f \Delta_h - \Delta_h \Delta_f)g = (\nabla_f \nabla_h - \nabla_h \nabla_f)\nabla_g V_a = (\nabla_f \nabla_h - \nabla_h \nabla_f)V_{a;g} =$$
$$= V_{a;g;h;f} - V_{a;g;f;h} = V_{c;g} R^c_{ahf} + V_{a;c} R^c_{ghf}$$

and

$$(\Delta_g \Delta_f - \Delta_f \Delta_g)h = (\nabla_g \nabla_f - \nabla_f \nabla_g)\nabla_h V_a = (\nabla_g \nabla_f - \nabla_f \nabla_g)V_{a;h} =$$
$$= V_{a;h;f;g} - V_{a;h;g;f} = V_{c;h} R^c_{afg} + V_{a;c} R^c_{hfg}$$

respectively. The left-hand side gives

$$(V_{a;f;g;h} - V_{a;g;f;h}) + (V_{a;h;f;g} - V_{a;f;h;g}) + (V_{a;g;h;f} - V_{a;h;g;f}) =$$
$$= (V_{a;f;g} - V_{a;g;f})_{;h} + (V_{a;h;f} - V_{a;f;h})_{;g} + (V_{a;g;h} - V_{a;h;g})_{;f} =$$
$$= (V_c R^c_{afg})_{;h} + (V_c R^c_{ahf})_{;g} + (V_c R^c_{agh})_{;f} =$$
$$= V_{c;h} R^c_{afg} + V_c R^c_{afg;h} + V_{c;g} R^c_{ahf} + V_c R^c_{ahf;g} + V_{c;f} R^c_{agh} + V_c R^c_{agh;f}.$$

The right-hand side gives

$$V_{c;h} R^c_{afg} + V_{a;c} R^c_{hfg} + V_{c;g} R^c_{ahf} + V_{a;c} R^c_{ghf} + V_{c;f} R^c_{agh} + V_{a;c} R^c_{fgh} =$$
$$= V_{c;h} R^c_{afg} + V_{c;g} R^c_{ahf} + V_{c;f} R^c_{agh} + V_{a;c} (R^c_{hfg} + R^c_{ghf} + R^c_{fgh}) =$$
$$= V_{c;h} R^c_{afg} + V_{c;g} R^c_{ahf} + V_{c;f} R^c_{agh}$$

as the term in $V_{a;c}$ vanishes due to the first Bianchi identity. The remaining right-hand side term cancels with an identical term on the left-hand side and we are left with

$$V_c (R^c_{afg;h} + R^c_{ahf;g} + R^c_{agh;f}) = 0.$$

The factor $V_c$ occurs throughout this equation and may be canceled out. We are left with the second Bianchi identity

$$R^c_{afg;h} + R^c_{ahf;g} + R^c_{agh;f} = 0.$$

Thus the first defining identity (the Jacobi identity) in (8) is analogous in form to the Bianchi identity in general relativity which is one set of gravitational field equations for the curvature tensor.

If $e$ stands for the metric tensor $g_{ab}$ and $\Delta_f$ for the covariant derivative $\nabla_f$, then the identity for the unit element is analogous to the important property that the metric is covariantly constant:

$$\Delta_f e \equiv \nabla_f g_{ab} = 0. \tag{10}$$

**The electromagnetic field**

If in equation (9), $(\Delta_h \Delta_g - \Delta_g \Delta_h) f$ is replaced by $(\partial_h \partial_g - \partial_g \partial_h) A_f$ where $A_f$ is the vector potential, cyclic permutations of indices produce analog expressions for the remaining terms of (9) and we obtain

$$(\partial_h \partial_g - \partial_g \partial_h) A_f + (\partial_f \partial_h - \partial_h \partial_f) A_g + (\partial_g \partial_f - \partial_f \partial_g) A_h = 0.$$

This equation can be rearranged as $\partial_h F_{gf} + \partial_f F_{hg} + \partial_g F_{fh} = 0$ which is the relativistic form of the homogeneous Maxwell equations where $F_{gh} = \partial_g A_h - \partial_h A_g$ is the electromagnetic tensor. Now, since $\alpha$ is an anti-symmetric product, $f \alpha g + g \alpha f = 0$ or in terms of the abstract derivation, $\Delta_g f + \Delta_f g = 0$. In terms of the vector potential this equation can be expressed as $\partial_g A_f + \partial_f A_g = 0$. If we subtract this equation from itself and rearrange terms we obtain $(\partial_g A_f - \partial_f A_g) + (\partial_f A_g - \partial_g A_f) = F_{gf} + F_{fg} = 0$. Thus the anti-symmetric nature of the $\alpha$ product is reflected in the antisymmetry of the electromagnetic tensor.

**Discussion**

Based on fundamental importance of the Hamiltonian in both mechanics, we did introduce an abstract derivation concept to generalize classical and quantum mechanics equations of motion to abstract equations of motion. The anti-symmetric product of the quantal algebra plays a key role in these abstract equations. When the defining identities of the quantal algebra are expressed in terms of the abstract derivation, certain concepts of the gauge theories (the algebraic identities and the homogeneous differential equations) are readily apparent.

A general terminology for gauge theories consists of seven concepts: the gauge group, the gauge field, the integrability conditions, the curvature, the algebraic identities, the homogeneous differential equations, and the inhomogeneous differential equations. Let us briefly illustrate these concepts for the gravitational and electromagnetic fields. For the gravitational field, the gauge group is the group of general coordinate transformations in a real four-dimensional Riemannian manifold with local Minkowski metric. The gauge field is the Christoffel symbol $\Gamma^{\sigma}_{\mu\nu}$.

The integrability condition is given by the relation $R^\sigma_{\rho\mu\nu} = 0$ where $R^\sigma_{\rho\mu\nu}$ is the Riemann curvature tensor. The curvature is defined exactly by the Riemann curvature tensor. The Riemann curvature tensor exhibits symmetry properties which can be read off directly from its definition:

$$R_{\sigma\rho\mu\nu} = -R_{\rho\sigma\mu\nu} = -R_{\sigma\rho\nu\mu} = R_{\mu\nu\sigma\rho}$$

$$R_{\sigma\rho\mu\nu} + R_{\sigma\nu\rho\mu} + R_{\sigma\mu\nu\rho} = 0$$

where $R_{\sigma\rho\mu\nu} = g_{\sigma\alpha} R^\alpha_{\rho\mu\nu}$ is the covariant form of the Riemann curvature tensor. These symmetry properties are referred to as algebraic identities. The second identity is referred to as the first Bianchi identity. Taking the covariant derivative of the Riemann tensor one obtains the homogeneous differential equations referred to as the second Bianchi identity:

$$R^\sigma_{\rho\mu\nu;\tau} + R^\sigma_{\rho\tau\mu;\nu} + R^\sigma_{\rho\nu\tau;\mu} = 0.$$

The Einstein equation $G_{\mu\nu} = 8\pi T_{\mu\nu}$ represents the inhomogeneous differential equations where $G_{\mu\nu} = R_{\mu\nu} - \frac{1}{2} g_{\mu\nu} R$ is the Einstein tensor and $T_{\mu\nu}$ is the stress-energy tensor. $R_{\mu\nu}$ is the Ricci tensor which is the first trace of the Riemann tensor. The second trace is the curvature scalar $R$.

Similarly, for the electromagnetic field, the gauge group is the unitary group of phase transformations $U(1)$. The gauge field is given by the four-potential $A_\mu(x)$. The integrability conditions are $\partial_\mu A_\nu - \partial_\nu A_\mu = 0$. The electromagnetic tensor $F_{\mu\nu} = \partial_\mu A_\nu - \partial_\nu A_\mu$ is the curvature, which thus appears as the analogue of Riemann's curvature tensor. The curvature tensor $F_{\mu\nu}$ is subject to only one algebraic identity. It is antisymmetry $F_{\mu\nu} + F_{\nu\mu} = 0$. The homogeneous differential equations are represented by the following differential identity

$$F_{\mu\nu,\lambda} + F_{\lambda\mu,\nu} + F_{\nu\lambda,\mu} = 0,$$

which is analogous to the second Bianchi identity in general relativity. The inhomogeneous differential equations are given by the source equation for the electromagnetic field $\partial_\rho F^{\mu\rho} = 4\pi J^\mu$.

In the gravitational field section we showed how the Jacobi identity of the quantal algebra is analogous in form to the Bianchi identities. Similarly, in the electromagnetic field section we showed how the Jacobi identity is analogous in form to the homogeneous differential equations. In addition, the anti-symmetric nature of the $\alpha$ product is reflected in the antisymmetry of the electromagnetic tensor. Perhaps it is not surprising why the identity for the unit element $e$ of the quantal algebra is analogous to the important property that the metric is covariantly constant. If $O_J$ is the centralizer of $J$, i.e. the set of all observables $f$ in $O$ such that $J \alpha f = 0$, where $J = \sqrt{-e}$, then $\{O_J, \sigma, \alpha, e\}$ is a quantal algebra [1]. $J$ plays a unique role in the algebra, and introduces relativity into the quantal framework. Specifically, once $J$ is selected, quantions are defined into a subspace of $SO(2,4)$, the centralizer space of $O_J(2,4)$. The centralizer reduces itself to a complex Minkowski space of dimensionality 8: $M_0(C) = M_0 \oplus iM_0$. Using the Leibnitz identity on $f \alpha (J \sigma J)$ we obtain the identity for the unit element ($e \alpha f = 0$), thus, $e$, in essence, introduces relativity into the quantal framework. $e$ is a unique element in the quantal algebra. Similarly, the metric tensor plays a unique role in general relativity. Equation 10 suggests they are analogous to each other.

Would the Leibnitz and Petersen identities of the quantal algebra (8) be somehow analogous to the inhomogeneous differential equations for the gravitational and electromagnetic fields? This is an open question.

## Conclusions

When the defining identities of the quantal algebra are expressed in terms of the abstract derivation, certain concepts of the gauge theories (the algebraic identities and the homogeneous differential equations) are readily apparent for the gravitational and electromagnetic fields.

**Acknowledgments**

The results presented in this paper are the outcome of independent research not supported by any institution or government grant.